\documentclass[nofootinbib,amsmath,amssymb,aps,floatfix,notitlepage,
longbibliography]{revtex4-2}
\pdfoutput=1

\usepackage[colorlinks,urlcolor=blue,citecolor=red]{hyperref}

\def\hhref#1{\href{http://arxiv.org/abs/#1}{arXiv:#1}} % in bibliography

\usepackage{geometry}                		% See geometry.pdf to learn the layout options. There are lots.
\usepackage{graphicx}				% Use pdf, png, jpg, or eps§ with pdflatex; use eps in DVI mode
\usepackage{amsmath,amssymb}

\begin{document}

\title{Instantons or Renormalons? A Comment on $\phi^4_{d=4}$ Theory in the MS Scheme}
\author{Gerald V. Dunne}
\affiliation{Department of Physics, University of Connecticut, Storrs CT 06269-3046, USA}
\author{Max Meynig}
\affiliation{Department of Physics, University of Connecticut, Storrs CT 06269-3046, USA}

%\date{\today}							% Activate to display a given date or no date

\begin{abstract}
We comment that the recent exact six-loop and seven-loop computations of renormalization group functions for the $O(N)$-symmetric four dimensional $\phi^4$ quantum field theory  show hints that the associated large order behavior is dominated by instantons rather than renormalons. This is consistent with a long-standing conjecture that renormalization group functions in the minimal subtraction (MS) renormalization scheme are not sensitive to renormalons.

\end{abstract}
\maketitle

\section{Introduction}

In quantum field theory (QFT) the two main sources of divergence of perturbation theory are identified as semiclassical ``instantons'' (more generally, ``saddles'') \cite{Lipatov:1976ny,zinn-book} or Feynman diagrammatic ``renormalons'' \cite{tHooft:1977xjm,Beneke:1998ui}. The divergence associated with instantons is typically combinatorial in nature \cite{Bender:1976ni,LeGuillou:1990nq}, related to the factorial proliferation of Feynman diagrams with the perturbative order. The divergence associated with renormalons is typically related to the momentum dependence of certain classes of iterated diagrammatic structures, such as bubble chains for example, and is closely related to the renormalization group and the operator product expansion \cite{Parisi:1978bj,Parisi:1978iq,Shifman:1978bx,Mueller:1992xz,Vainshtein:1994qq}.
These divergences of perturbation theory appear as singularities in the Borel plane of the corresponding Borel transform of the perturbative expansion of the quantity that is being computed. The dominant large-order growth of the perturbative coefficients corresponds to the dominant non-perturbative effects, and to the Borel singularity(ies) closest to the origin of the Borel plane.
The generic form of the {\it leading} large-order ($k\to\infty$) growth  of the perturbative coefficients $c_k$ has the 
canonical {\it power-times-factorial} form \cite{LeGuillou:1990nq}:
\begin{eqnarray}
c_k\sim {\mathcal S} \, a^k\, \Gamma(k+1+b) \qquad , \quad k\to\infty
\label{eq:bwl}
\end{eqnarray}
The three parameters $a$, $b$, and ${\mathcal S}$ have the following physical significance:
\begin{eqnarray}
\frac{1}{a} &\leftrightarrow& \rm{location\, of\, the\, leading\, Borel\, singularity} \\
b &\leftrightarrow& \rm{nature\, of\, the\, leading\, Borel\, singularity} \\
{\mathcal S} &\leftrightarrow& \rm{strength\, of\, the\, leading\, Borel\, singularity}
\label{eq:meaning}
\end{eqnarray}
While these correspondences are well understood in quantum mechanics, matrix models and certain special quantum field theories, it is much more difficult to make high-order perturbative computations of Green's function or renormalization group (RG) functions in non-trivial QFTs.
For example, the beta functions in QED \cite{Kataev:2012rf} and QCD \cite{Herzog:2017ohr} are currently known to 5 loop order, as is the QED anomalous magnetic moment of the electron \cite{Aoyama:2012wj,aoyama}, each of which constitutes a heroic {\it tour de force} computation. The anomalous magnetic moment of the muon is a question of great current interest \cite{Aoyama:2020ynm}.

Recently there has been dramatic progress in the understanding of QFT perturbative amplitudes \cite{Bern:1991aq,Kreimer:1997dp,Arkani-Hamed:2013jha,Bern:2010ue}, leading to new methods for high-order computations. For example, following the 5-loop analysis of $\phi^4$ theory  in the early 1980s \cite{Gorishnii:1983gp,Gorishnii:1983jr,Kleinert:1991rg}, the 6-loop \cite{panzer} and 7-loop \cite{schnetz} computations of the beta function and anomalous dimensions (in MS scheme) have been done in quick succession in recent years. These computations have been inspired and enabled by deep ideas from graph theory, number theory and Hopf algebras \cite{Broadhurst:1995km,Panzer:2016snt,Schnetz:2016fhy}.

Here we ask the following simple question:
\begin{quote}
{\it Do the exact results of \cite{panzer,schnetz} contain enough perturbative data to be able to see hints of large-order growth and associated non-perturbative effects in $\phi^4$ theory in 4 dimensions, and to distinguish between instanton or renormalon effects?}
\end{quote}

We suggest that the answer is ``yes'', and that the results so far appear to favor the instanton description. The idea is extremely simple:  the results of  \cite{panzer,schnetz} include the full $N$-dependence for the $O(N)$ symmetric $\phi^4$ model, and the instanton and renormalon predictions for the $N$-dependence of the large-order growth parameters  $a$, $b$, and ${\mathcal S}$ appearing in (\ref{eq:bwl}) are very different.

The instanton analysis \cite{Lipatov:1976ny,Brezin:1976vw,McKane:1978md,McKane:2018ocs} predicts that the leading large-order growth of the coefficients $\beta_k$ of the beta function is:
\begin{eqnarray}
{\rm instantons:}\quad \beta^{\rm inst}_k \sim (-1)^k  {\mathcal S}_{\rm inst}\, \Gamma\left(k+4+\frac{N}{2}\right)\qquad, \quad k\to \infty
\label{eq:betak-inst}
\end{eqnarray}
As is conventional \cite{panzer}, we have absorbed a factor of the single-instanton action (which is independent of $N$) into the coupling.\footnote{Thus, the Lagrangian for the $\mathcal{O}(N)$ symmetric field $\phi$ in $D=4-2\epsilon$ dimensions is
\begin{equation}
    \mathcal{L} = \frac{1}{2}\, m^2 \, Z_1\, \phi^2 +\frac{1}{2}\,Z_2\, (\partial \phi )^2 + \frac{16 \pi^2}{4! } Z_4\, g\, \mu^{2\epsilon}\,\phi^4
    \label{eq:Lagrangian}
\end{equation}}
The overall constant ${\mathcal S}_{\rm inst}$ is  known as a function of $N$, but this will not enter our argument here.

The renormalon analysis \cite{Parisi:1978bj,Parisi:1978iq,deCalan:1981szv,Magnen:1987gy,david,David:1988bi} leads to the following quite different prediction:
\begin{eqnarray}
{\rm renormalons:}\quad \beta^{\rm ren}_k \sim {\mathcal S}_{\rm ren}\, \left(\frac{\beta_2(N)}{2}\right)^k \, \Gamma\left(k+1+\frac{2\beta_3(N)}{(\beta_2(N))^2}\right)\qquad, \quad k\to \infty
\label{eq:betak-ren}
\end{eqnarray}
where $\beta_2(N)=\frac{N+8}{3}$ and $\beta_3(N)=\frac{3N+14}{3}$ are the first two non-trivial beta function coefficients. 
We can summarize these different predictions for the large-order growth parameters $a$ and $b$ in (\ref{eq:bwl}) as:
\begin{eqnarray}
a_{\rm inst}(N) =-1\qquad &;& \qquad b_{\rm inst}(N)=3+\frac{N}{2}
\label{eq:ab-inst} \\
a_{\rm ren}(N) = \frac{(N+8)}{6}\qquad &;& \qquad b_{\rm ren}(N)=\frac{6(3N+14)}{(N+8)^2}
%\frac{2\beta_3(N)}{(\beta_2(N))^2}
\label{eq:ab-ren}
\end{eqnarray}
We note that there is an old conjecture that the renormalization group functions in MS scheme are not sensitive to renormalons -- see the discussion in \cite{McKane:2018ocs} and comments in \cite{david}.

It is possible to selectively probe the large-order growth parameters $a$ and $b$ in (\ref{eq:bwl}) as follows. For coefficients with this {\it factorial-times-power} large-order growth, the ratio-of-ratios $c_{k+1}c_{k-1}/c_k^2$ should tend to $1$ at large order as follows:\footnote{This is a common indicator \cite{guttmann,Ellis:1995jv}.}
\begin{eqnarray}
\frac{c_{k+1}c_{k-1}}{c_k^2} \sim 1+O\left(\frac{1}{k}\right) \qquad, \quad k\to\infty
\label{eq:r2}
\end{eqnarray}
The subleading $O\left(\frac{1}{k}\right)$ correction term is directly sensitive to the factorial growth and the associated large-order growth parameter $b$ (the $a$ dependence clearly cancels):
\begin{eqnarray}
(k+b)\left(\frac{c_{k+1}c_{k-1}}{c_k^2} - 1 \right) \sim 1+O\left(\frac{1}{k}\right) \qquad, \quad k\to\infty
\label{eq:ratioratio}
\end{eqnarray}
If the parameter $b$ has been determined, the simple ratio $c_{k+1}/c_k$ can be used to determine the other large-order growth parameter, $a$:
\begin{eqnarray}
\frac{1}{a} \frac{1}{(k+1+b)}\left(\frac{c_{k+1}}{c_k}\right) \sim 1+O\left(\frac{1}{k}\right) \qquad, \quad k\to\infty
\label{eq:ratio}
\end{eqnarray}
Therefore, the combinations of coefficients on the LHS of  (\ref{eq:r2}), (\ref{eq:ratioratio}) and (\ref{eq:ratio}) should each tend to $1$ at large perturbative order $k$.

\section{Perturbative Expansion of the Beta Function}

To fix notation, we list the first few terms of the perturbative RG beta function, computed in the MS renormalization scheme \cite{panzer,schnetz}. 
The first few terms for the beta function expansion are:
\begin{eqnarray}
\beta(g, N, \epsilon)&:=&\sum_{k=0}^\infty\beta_k(N)\, g^k
\label{eq:betakn}\\
&=& -2 \epsilon g+\left(\frac{8+N}{3}\right)g^2 -\left(\frac{14+3N}{3}\right)g^3  \nonumber\\
&&+\left(\left(\frac{88 \zeta(3)}{9}+\frac{370}{27}\right)+\left(\frac{20 \zeta(3)}{9}+\frac{461}{108}\right) N+\frac{11 N^{2}}{72}\right) g^4  \nonumber\\
&&-\left(\left(\frac{24581}{486}+\frac{4664 \zeta(3)}{81}+\frac{2480 \zeta(5)}{27} -\frac{176 \pi^{4}}{1215}\right)+\left(\frac{10057}{486}+\frac{1528 \zeta(3)}{81}+\frac{2200 \zeta(5)}{81} -\frac{62 \pi^{4}}{1215}\right) N \right.\nonumber \\
&&\left. +\left(\frac{395}{243}+\frac{14 \zeta(3)}{9}+\frac{80 \zeta(5)}{81}-\frac{\pi^{4}}{243}\right) N^{2}- \frac{5 N^{3}}{3888} \right) g^{5} +\dots
\label{eq:beta7}
\end{eqnarray}
The terms to 6-loop are in \cite{panzer} and to 7-loop in \cite{schnetz}.

An approximation to the beta function based on the primitive diagrams (those without subdivergences) has been computed to 11 loop order \cite{panzer,hepp,erik} (presented here in 4 dimensions where $\epsilon=0$):
  \begin{equation}   \label{eq:beta-primitive}
  \begin{aligned}
    &\beta_{\rm primitive}(g, N)=   \frac{1}{3} g^2 \left( 8+N\right)- 0\cdot g^3 +g^4 (11.7534\, +2.67124 N)-g^5 \left(95.2437\, +28.1635 N+1.02413 N^2\right)\\
    &+g^6\left(1226.29\, +438.768 N+33.1118 N^2\right)-g^7 \left(16490.3\, +6872.79 N+751.561 N^2+16.0652 N^3\right)\\
    &+g^8 \left(240539.\, +113676.0 \, N+16034.7 N^2+672.775 N^3+2.59286 N^4\right) -g^9 \big(3.73942\times 10^6 \\ 
    &+1.96561\times 10^6 N +335648.0 \, N^2+20839.2 N^3+337.374 N^4\big)+g^{10} \big(6.14646\times 10^7 \\
    &+3.54108\times 10^7 N +7.03792\times 10^6 N^2+572454.0 \, N^3+16798.5 N^4+88.9656 N^5\big)\\
    &-g^{11} \big(1.06184\times 10^9+6.6272\times 10^8 N+1.49249\times 10^8 N^2+1.48812\times 10^7 N^3+634038.0 \, N^4\\
    &+8689.46 N^5+9.51934 N^6\big)+g^{12} \big(1.92531\times 10^{10}+1.28903\times 10^{10} N+3.22426\times 10^9 N^2 \\
    &+3.77652\times 10^8 N^3+2.09798\times 10^7 N^4+481120.0 \, N^5+2820.02 N^6\big)
  \end{aligned}
\end{equation}
The primitive diagrams constitute the dominant  fraction of diagrams at large order, and their contribution to the RG functions is scheme independent \cite{panzer}.

We use the results of \cite{panzer,schnetz} and (\ref{eq:beta-primitive}) to study the coefficient combinations in (\ref{eq:r2}), (\ref{eq:ratioratio}) and (\ref{eq:ratio}) using the instanton and renormalon predictions in (\ref{eq:ab-inst})--(\ref{eq:ab-ren}) for the large-order growth parameters $a$ and $b$, including their $N$ dependence. We first define
\begin{eqnarray}
\delta_k(N):=\frac{\beta_{k+1}(N) \beta_{k-1}(N)}{\beta_k^2(N)} 
\label{eq:delta-kN}
\end{eqnarray}
\begin{figure}[h!]
\includegraphics[width=7cm]{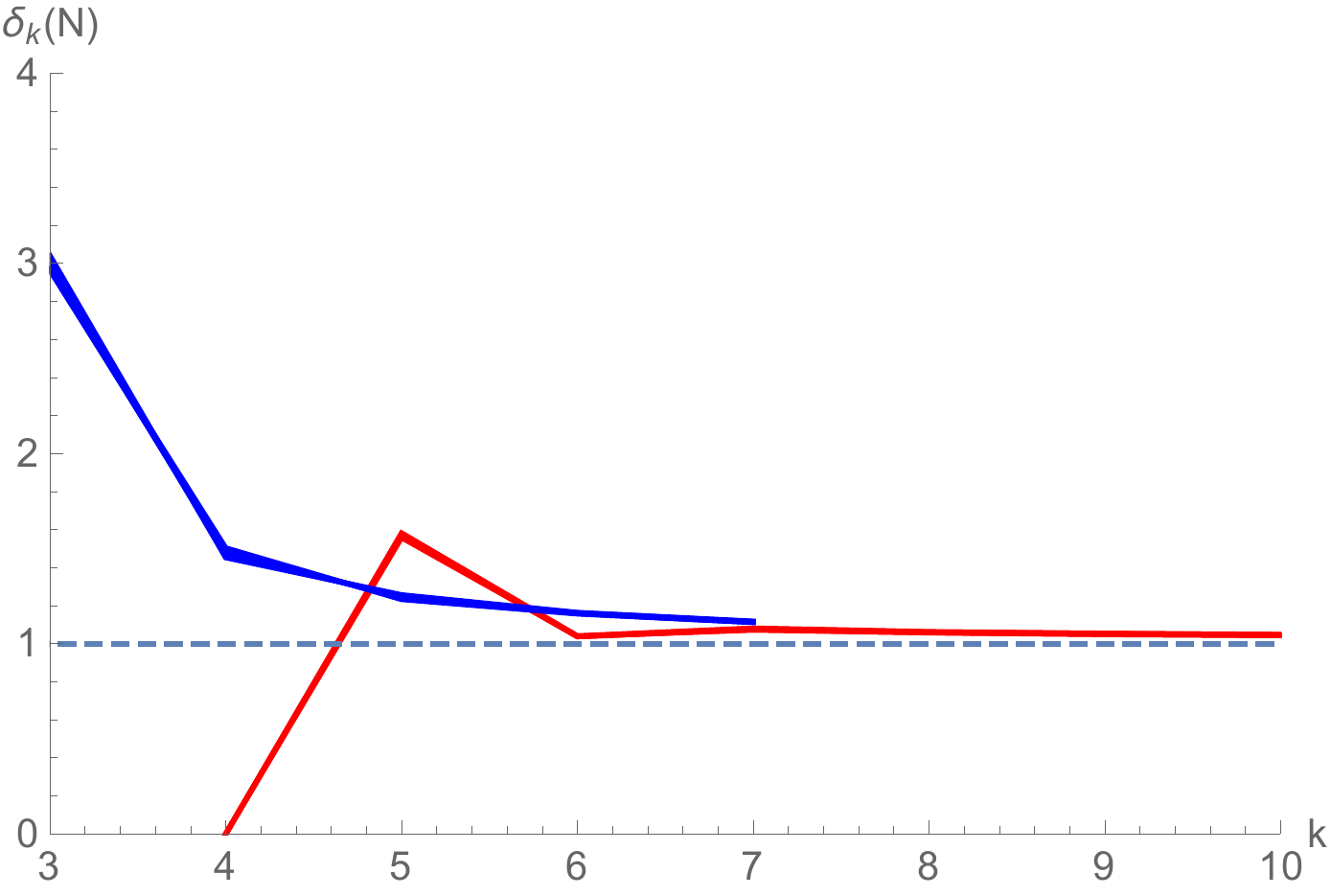}
\caption{Plots of the ratio $\delta_k(N)$ defined in (\ref{eq:delta-kN}) as a function of the perturbative order $k$,  for $N=1,2, ..., 5$. The dashed horizontal line $\delta_k=1$ is the predicted large $k$ limit. The blue  curves use the exact 7-loop beta function coefficients $\beta_k(N)$ in (\ref{eq:beta7}) from \cite{panzer,schnetz}, while the red curves use the approximate 11-loop primitive graph beta function coefficients \cite{erik} in (\ref{eq:beta-primitive}).}
\label{fig:raw-ratio}
\end{figure}
If the beta function coefficients $\beta_k(N)$ follow the large-order growth in (\ref{eq:bwl}) then $\delta_k(N)$ should tend to $1$,
 independent of the values of the large order growth parameters $a$, $b$ and ${\mathcal S}$ in (\ref{eq:bwl}).\footnote{Clearly, $N$ cannot be too large, or the formal perturbative expansion must be re-organized.} 
 Figure \ref{fig:raw-ratio} plots $\delta_k(N)$ as a function of perturbative order $k$, for $N=1, 2, .., 5$. This figure suggests that  both the exact coefficients to 7-loop order and  the (approximate) primitive-graph results to 11-loop order are consistent with the form of the large-order growth in (\ref{eq:bwl}).
\begin{figure}[h!]
\includegraphics[width=7cm]{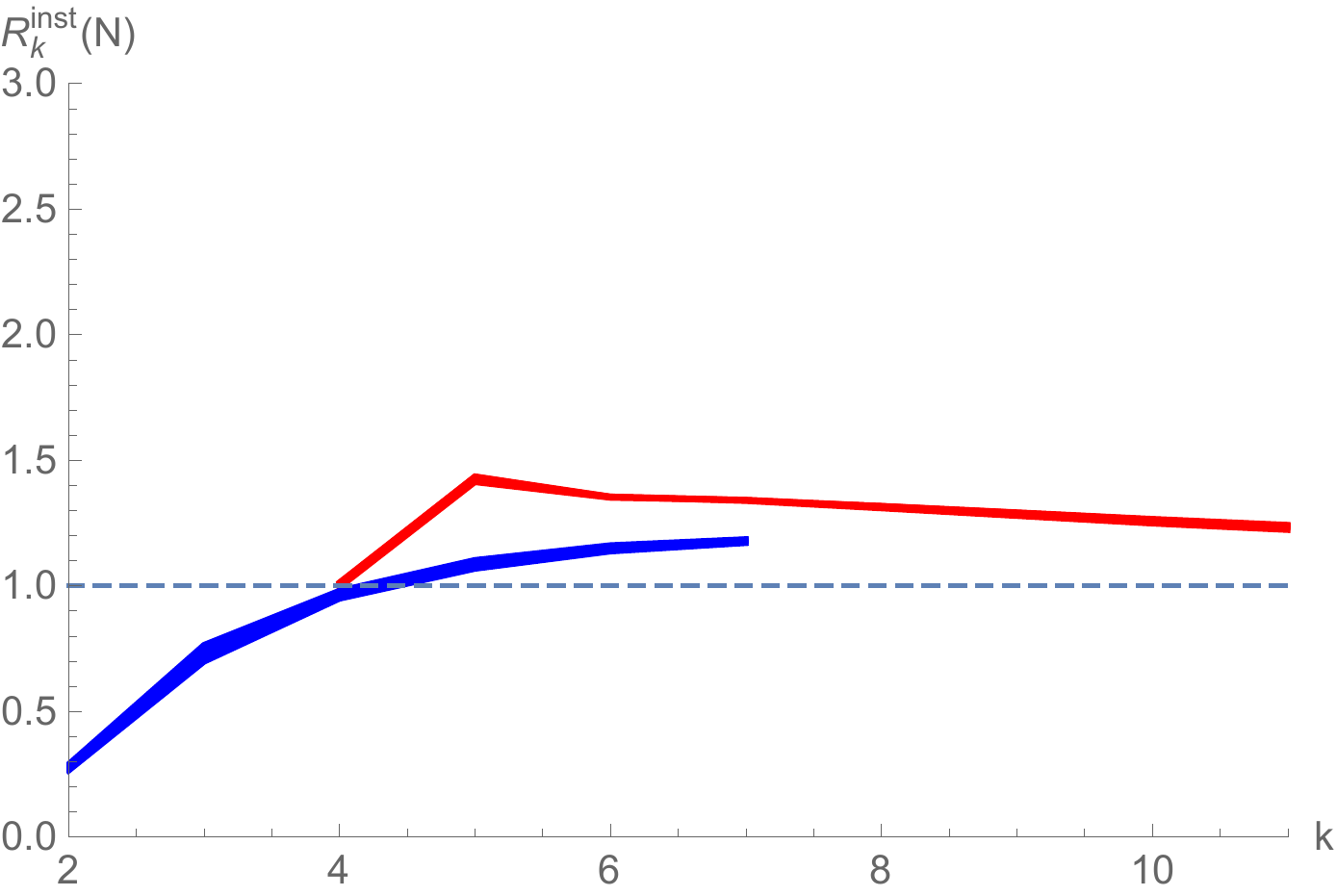}\quad
\includegraphics[width=7cm]{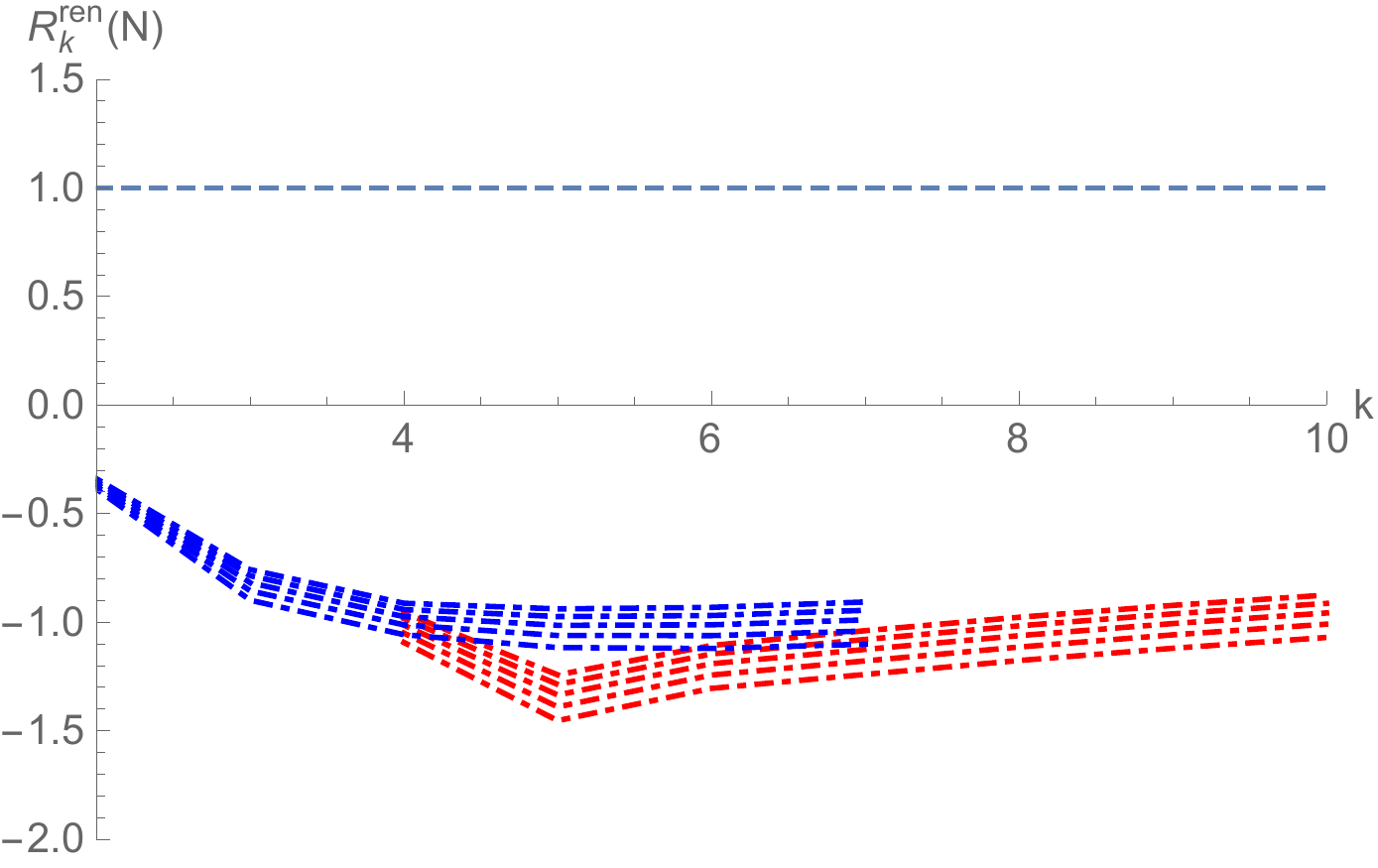}
\caption{Left-hand Figure: Plots of the ratio ${\mathcal R}_k^{\rm inst}(N)$ defined in (\ref{eq:R-inst}), based on the instanton large-order growth parameters in (\ref{eq:ab-inst}), as a function of the perturbative order $k$, for $N=1,2, ..., 5$. The blue curves use the exact 7-loop beta function coefficients $\beta_k(N)$ in (\ref{eq:beta7}) from \cite{panzer,schnetz}, while the red curves use the approximate 11-loop primitive graph beta function coefficients \cite{erik} in (\ref{eq:beta-primitive}). The dashed horizontal line ${\mathcal R}_k=1$ is the predicted large $k$ limit. 
Right-hand Figure:  Plots of the ratio ${\mathcal R}_k^{\rm ren}(N)$ defined in (\ref{eq:R-ren}), based on the renormalon large-order growth parameters in (\ref{eq:ab-ren}).  The color-coding is the same as in the left-hand Figure. }
\label{fig:beta-ratio-inst-plot}
\end{figure}

To probe this more precisely we define the ratio of successive beta function coefficients, normalized as in (\ref{eq:ratio}):
\begin{eqnarray}
{\mathcal R}_k^{\rm inst}(N)&:=&\frac{1}{a^{\rm inst}(N)\, (k+1+b^{\rm inst}(N))} \frac{\beta_{k+1}(N)}{\beta_k(N)}  = \frac{-1} {\left(k+4+\frac{N}{2}\right)}   \frac{\beta_{k+1}(N)}{\beta_k(N)}
\label{eq:R-inst}
\\
{\mathcal R}_k^{\rm ren}(N)&:=&\frac{1}{a^{\rm ren}(N)\, (k +1+b^{\rm ren}(N))} \frac{\beta_{k+1}(N)}{\beta_k(N)}  =\frac{6}{(N+8)\left(k+1+\frac{6(3N+14)}{(N+8)^2}\right)} \frac{\beta_{k+1}(N)}{\beta_k(N)}
\label{eq:R-ren}
\end{eqnarray}
For consistency, each of these should tend to $1$ at large perturbative order $k$. From Figure \ref{fig:beta-ratio-inst-plot}  we see that the instanton parameters $a_{\rm inst}(N)$ and $b_{\rm inst}(N)$ in (\ref{eq:ab-inst}) are clearly favored over the renormalon parameters in (\ref{eq:ab-ren}), both because of the sign of $a$ and because of the $N$ dependence.

\begin{figure}[h!]
\includegraphics[width=7cm]{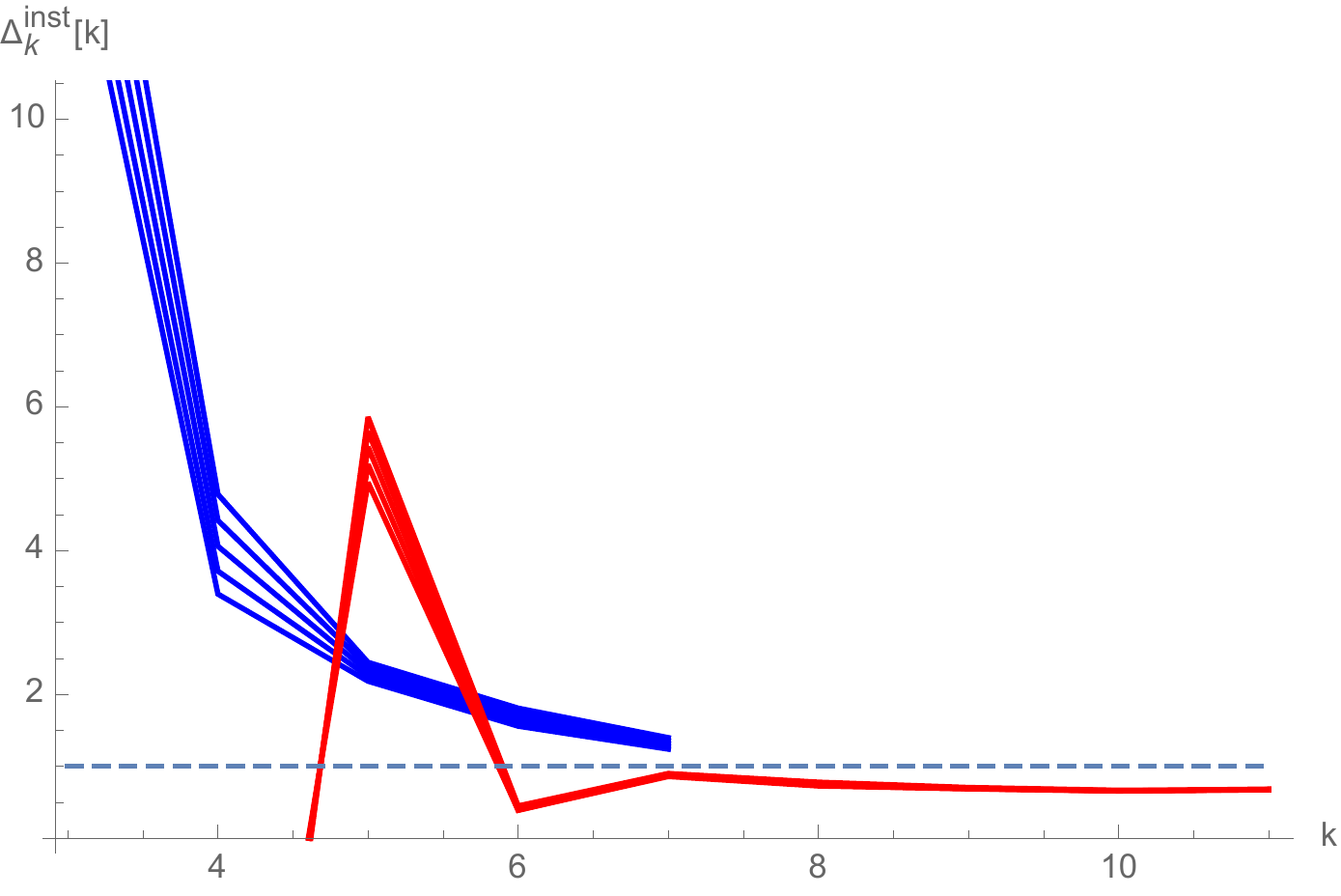}
\quad
\includegraphics[width=7cm]{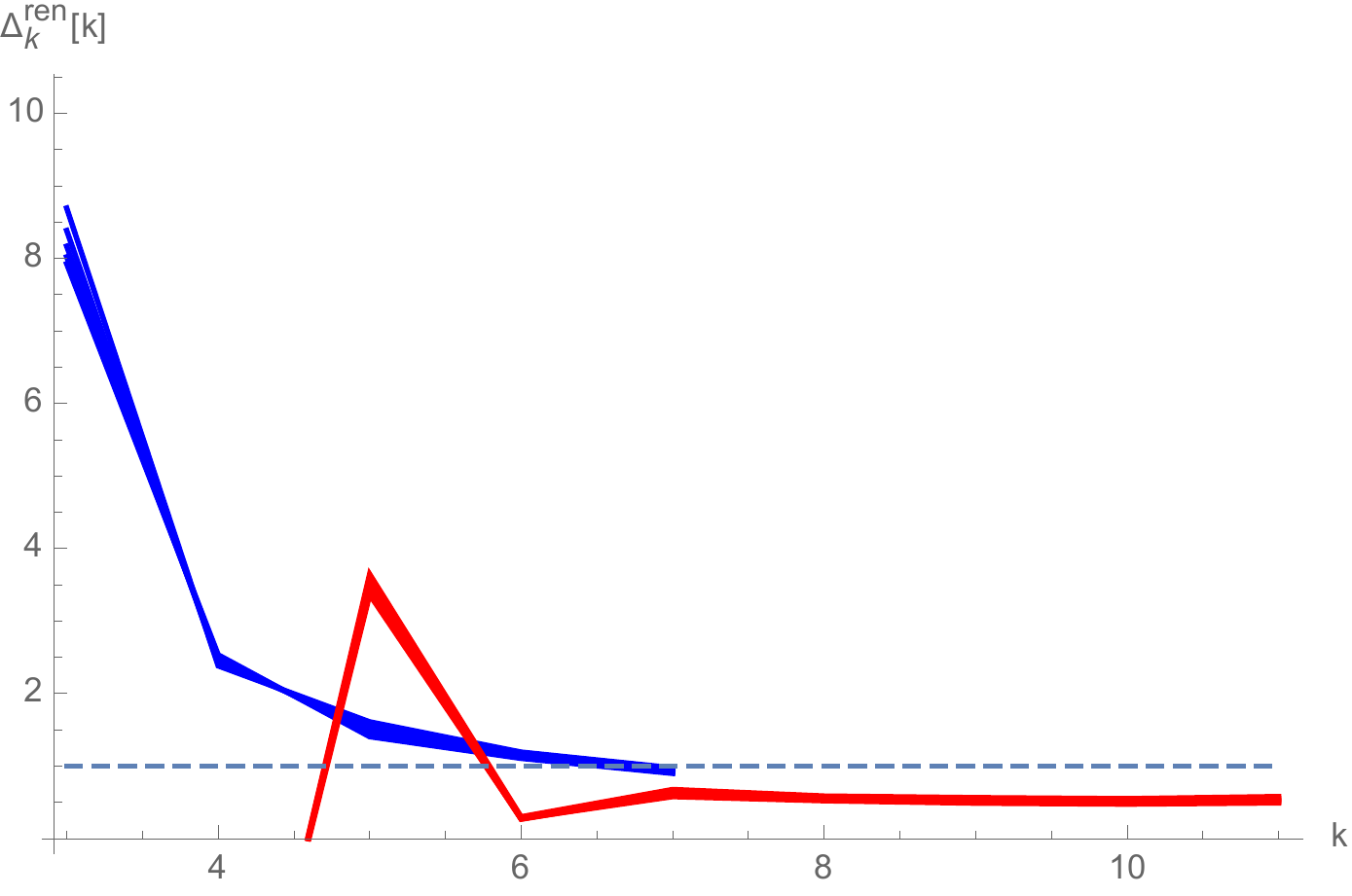}
\caption{Left-hand Figure: Plots of the ratio $\Delta_k^{\rm inst}(N)$ defined in (\ref{eq:D-inst}), based on the instanton large-order growth parameters in (\ref{eq:ab-inst}), as a function of the perturbative order $k$, for $N=1,2, ..., 5$. The blue curves use the exact 7-loop beta function coefficients $\beta_k(N)$ in (\ref{eq:beta7}) from \cite{panzer,schnetz}, while the red curves use the approximate 11-loop primitive graph beta function coefficients \cite{erik} in (\ref{eq:beta-primitive}). The dashed horizontal line $\Delta_k=1$ is the predicted large $k$ limit. 
Right-hand Figure:  Plots of the ratio $\Delta_k^{\rm ren}(N)$ defined in (\ref{eq:D-ren}), based on the renormalon large-order growth parameters in (\ref{eq:ab-ren}).  The color-coding is the same as in the left-hand Figure.}
\label{fig:beta-delta-inst-plot}
\end{figure}

Finally, we probe the large-order growth parameter $b$ by plotting the large-order growth of the following combinations, normalized as in (\ref{eq:ratioratio}):
\begin{eqnarray}
\Delta_k^{\rm inst}(N)&:=&\left(k+3+\frac{N}{2}\right)\left(\frac{\beta_{k+1}(N) \beta_{k-1}(N)}{\beta_k^2(N)} - 1 \right)
\label{eq:D-inst}
\\
\Delta_k^{\rm ren}(N)&:=&\left(k+\frac{6(3N+14)}{(N+8)^2}\right)\left(\frac{\beta_{k+1}(N) \beta_{k-1}(N)}{\beta_k^2(N)} - 1 \right)
\label{eq:D-ren}
\end{eqnarray}
Again, for consistency each of these should tend to $1$. Figure \ref{fig:beta-delta-inst-plot} slightly favors the instanton parameters over the renormalon ones, but the difference is not as conclusive as in Figure \ref{fig:beta-ratio-inst-plot}.

\section{Epsilon Expansion: Critical Coupling and Correction to Scaling Exponent}

Given the perturbative expansion (\ref{eq:betakn}) of the beta function, we use straightforward series inversion to compute the epsilon expansion of the critical coupling, defined via  \cite{zinn-book}:
   \begin{equation}\label{eq:CritCouplingDef}
       \beta (g_{\rm crit}(\epsilon,N),N,\epsilon ) = 0 
   \end{equation}
The first few terms are
    \begin{eqnarray}
        g_{\rm crit}(\epsilon,N)&:=&\sum_{k=1}^\infty g_k(N)\, \epsilon^k
        \label{eq:gcrit}
    \\
        &=&  \frac{6 \epsilon }{N+8} +\frac{36 (3 N+14) \epsilon ^2}{(N+8)^3} 
            \nonumber\\
        &&+ \frac{3 \epsilon ^3 \left(-96 \zeta (3) (N+8) (5 N+22)+11 N (N (10-3 N)+160)+4544\right)}{(N+8)^5} +\dots
    \label{eq:CritCoupling}
\end{eqnarray}
The epsilon expansion inherits a related factorial divergence from the perturbative results computed for the RG functions in dimension $D=4-2\epsilon$: 
\begin{eqnarray}
g_k(N)\sim \left(-\frac{2}{\beta_2(N)}\right)^k \Gamma\left(k+\frac{N}{2}+ 4 \right)
 \qquad,\qquad k\to\infty
\label{eq:gc-large-order}
\end{eqnarray}
The large order behavior of $g_k(N)$ is controlled by the same offset as for the beta function coefficients $\beta_k(N)$, but  the ``action" is rescaled by $2/\beta_2(N)$, as can be seen from the first term in (\ref{eq:beta7}).
We have adopted the instanton large-order growth parameters, since these are favored by the beta function analysis above.

The correction to scaling exponent is defined as  \cite{zinn-book}
   \begin{equation}\label{eq:OmegaDef}
       \omega(\epsilon,N) \equiv \left[\frac{d}{dg} \beta(g,N,\epsilon)\right]_{g=g_{\rm crit}(\epsilon,N)} .
   \end{equation}
Thus, the epsilon expansion of $\omega(\epsilon,N)$ is obtained by series composition.  The first few terms are
\begin{eqnarray}
       \omega(\epsilon,N)&:=&\sum_{k=1}^\infty \omega_k(N)\, \epsilon^k
      \label{eq:w}\\
       &=&2 \epsilon -\frac{12 (3 N+14) \epsilon ^2}{(N+8)^2}  \nonumber\\
       &&+ \epsilon ^3 \left(\frac{192 \zeta (3) (5 N+22)}{(N+8)^3}+\frac{2 (N (N (33 N+538)+4288)+9568)}{(N+8)^4}\right) +\dots
\label{eq:Omega}
\end{eqnarray}
The large order behavior of coefficients $\omega_k(N)$ in (\ref{eq:w})  is more complicated because it is influenced by two series, the perturbative series of $\beta(g,N,\epsilon)$ and the epsilon expansion of $ g_{\rm crit}(\epsilon,N)$, each of which has large-order behavior of the form (\ref{eq:bwl}), but with different parameters $a$ and $b$. The derivative of the beta function causes a shift in the offset parameter $b$, while the $a$ parameter of the critical coupling dominates over that of the beta function.
The leading large order behavior is (see also \cite{panzer})
\begin{equation}
    \omega_k(N) \sim \left(-  \frac{2}{\beta_2} \right)^{k+\frac{N}{2}+6} \Gamma \left(k+\frac{N}{2}+6 \right) + \dots
    \qquad,\qquad k\to\infty
    \label{eq:omega-large}
\end{equation}
assuming that the large order behavior of the beta function is controlled by instanton arguments.

To compare this prediction (\ref{eq:omega-large}) with the perturbative results of \cite{panzer,schnetz}, we form the corresponding combinations of coefficients on the left-hand-sides of (\ref{eq:r2}), (\ref{eq:ratioratio}) and (\ref{eq:ratio}), analogous to the expressions (\ref{eq:delta-kN}), (\ref{eq:R-inst}) and (\ref{eq:D-inst}) for the beta function coefficients:
\begin{figure}[h!]
\includegraphics[width=7cm]{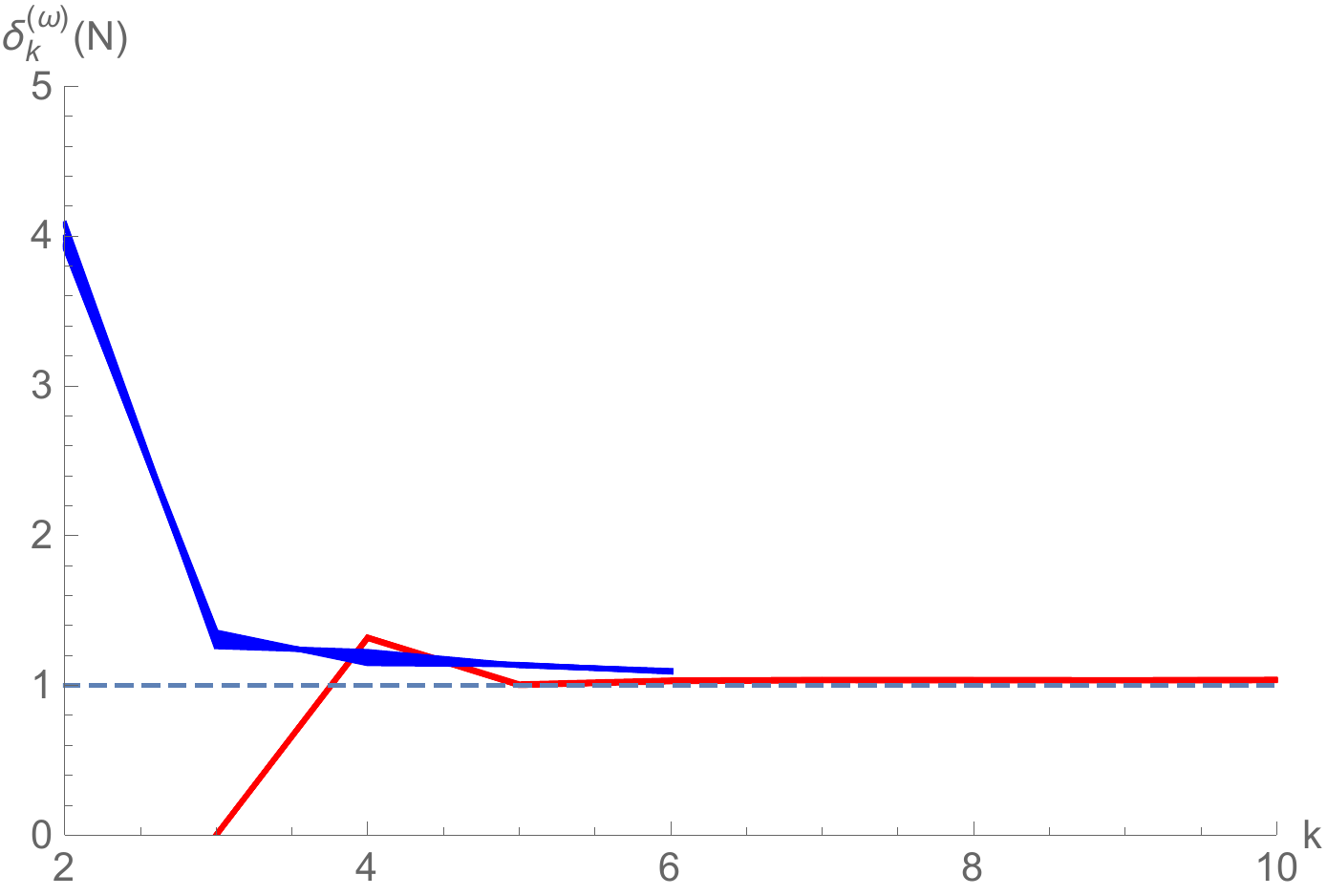}
\caption{Plot of the ratio $\delta_k^{(\omega)}(N)$ defined in (\ref{eq:delta-kN-omega}) as a function of the perturbative order $k$,  for $N=1,2, ..., 5$. The dashed horizontal line $\delta_k=1$ is the predicted large $k$ limit. The blue curves use the exact 7-loop beta function coefficients $\beta_k(N)$ in (\ref{eq:beta7}) from \cite{panzer,schnetz}, while the red curves use the approximate 11-loop primitive graph beta function coefficients \cite{erik} in (\ref{eq:beta-primitive}).}
\label{fig:omega-ratio-plot}
\end{figure}
\begin{figure}[h!]
\includegraphics[width=7cm]{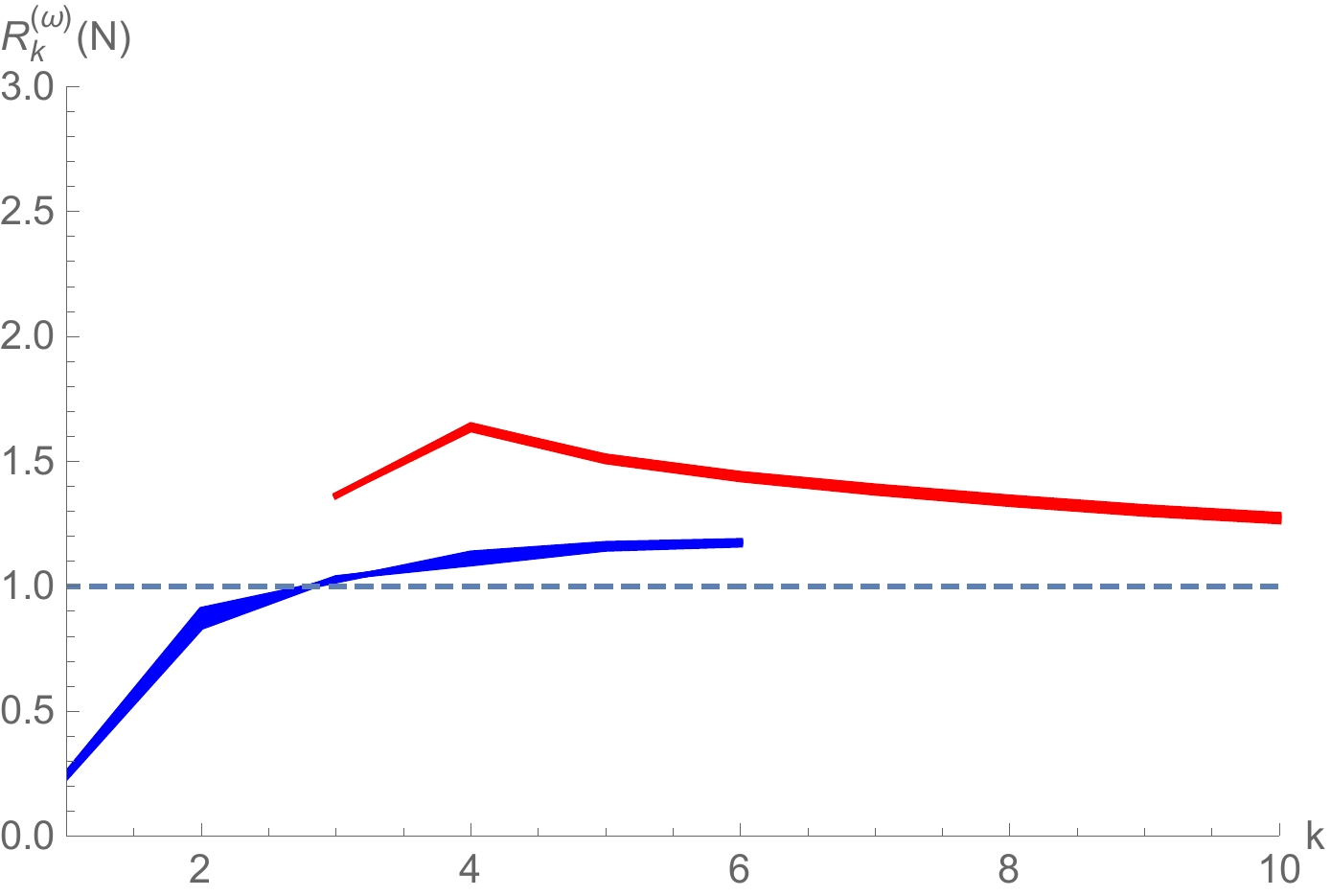}
\quad
\includegraphics[width=7cm]{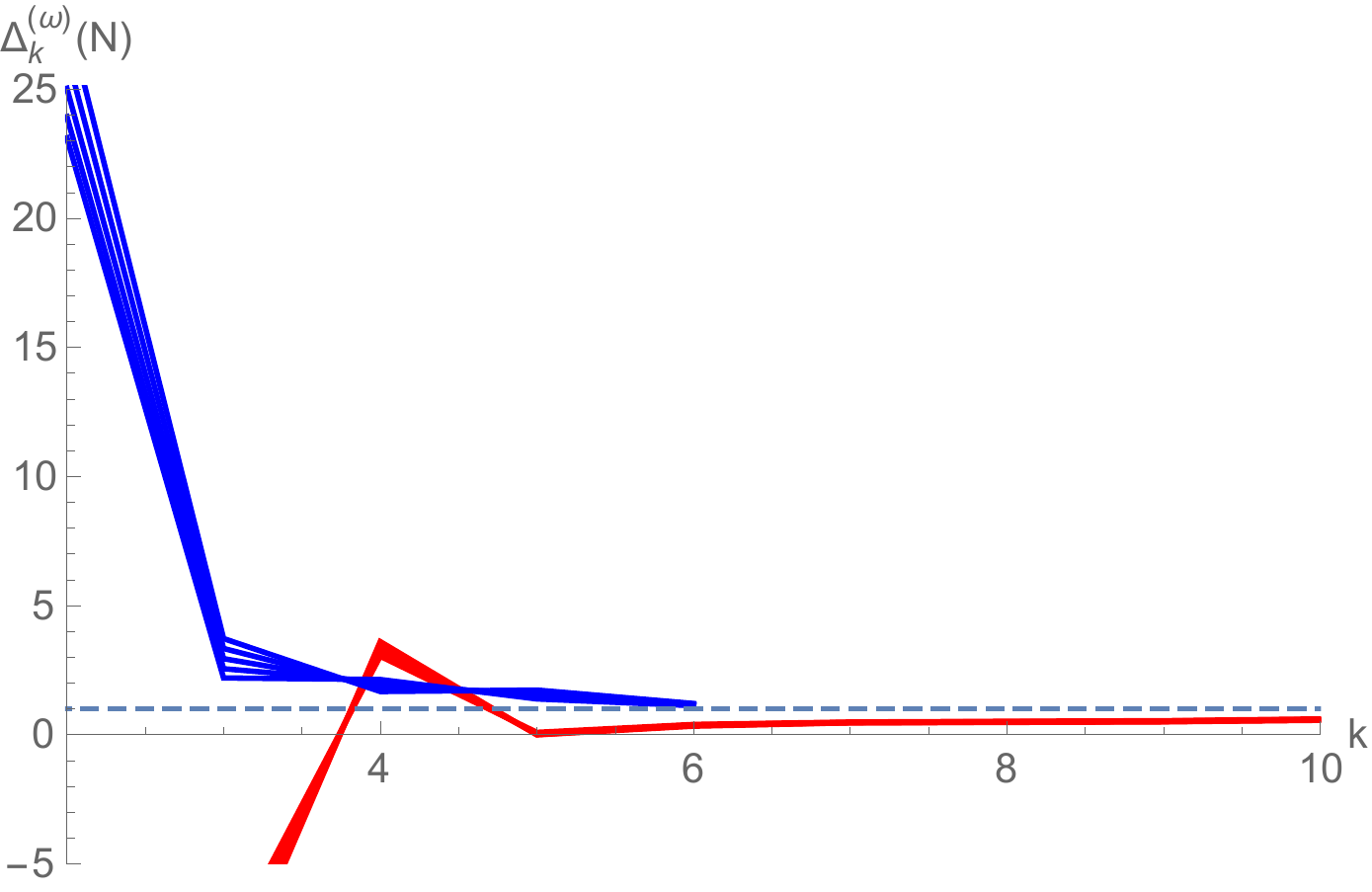}
\caption{Plots of the ratio ${\mathcal R}_k^{(\omega)}(N)$ defined in (\ref{eq:R-omega}) [left hand figure] and the ratio $\Delta_k^{(\omega)}(N)$ defined in (\ref{eq:D-omega}) [right hand figure], as  functions of the perturbative order $k$,  for $N=1,2, ..., 5$. The dashed horizontal lines ${\mathcal R}_k=1$ and $\Delta_k=1$ are the predicted large $k$ limit. The blue curves use the exact 7-loop beta function coefficients $\beta_k(N)$ in (\ref{eq:beta7}) from \cite{panzer,schnetz}, while the red curves use the approximate 11-loop primitive graph beta function coefficients \cite{erik} in (\ref{eq:beta-primitive}).}
\label{fig:omega-R-D-plot}
\end{figure}
\begin{eqnarray}
\delta_k^{(\omega)}(N)&:=&\frac{\omega_{k+1}(N)\, \omega_{k-1}(N)}{\omega_k^2(N)} 
\label{eq:delta-kN-omega} \\
{\mathcal R}_k^{(\omega)}(N)&:=& \frac{-(N+8)} {6 \left(k+6+\frac{N}{2}\right)}   \frac{\omega_{k+1}(N)}{\omega_k(N)} 
\label{eq:R-omega} \\
\Delta_k^{(\omega)}(N)&:=&\left(k+5+\frac{N}{2}\right)\left(\frac{\omega_{k+1}(N) \,\omega_{k-1}(N)}{\omega_k^2(N)} - 1 \right)
\label{eq:D-omega}
\end{eqnarray}
If the epsilon expansion coefficients $\omega_k(N)$ in (\ref{eq:Omega}) follow the large-order growth in (\ref{eq:omega-large}) then these combinations should each tend to $1$, for all $N$.
Figures \ref{fig:omega-ratio-plot} and \ref{fig:omega-R-D-plot} show good agreement between the predicted large-order behavior  (\ref{eq:omega-large}) and the perturbative results of \cite{panzer,schnetz}, similar to the large-order behavior of the beta function coefficients.

\section{Quantum Mechanical Model}

It is a simple but instructive exercise to compare these results for the 4 dimensional $O(N)$ symmetric $\phi^4$ QFT with its $1$ dimensional counterpart, the $O(N)$ symmetric quartic anharmonic oscillator, whose large-order growth for energy levels was analyzed in \cite{Banks:1973ps}. The ground state energy has a perturbative expansion
\begin{equation}
    E(\lambda, N) = \sum_{k=0}^\infty E_k(N) \,\lambda^k  
    \label{eq:ground}
\end{equation}
where $E_0(N)=\frac{N}{2}$, and the leading large order growth is \cite{Banks:1973ps}
\begin{eqnarray}
E_k(N) \sim - \frac{6^{N/2}}{\pi \Gamma\left(\frac{N}{2}\right)} \, (-3)^k \, \Gamma \left(k + \frac{N}{2} \right)\qquad, \quad k\to \infty
\label{eq:ground-large}
\end{eqnarray}
\begin{figure}[h!]
 \includegraphics[width=7cm]{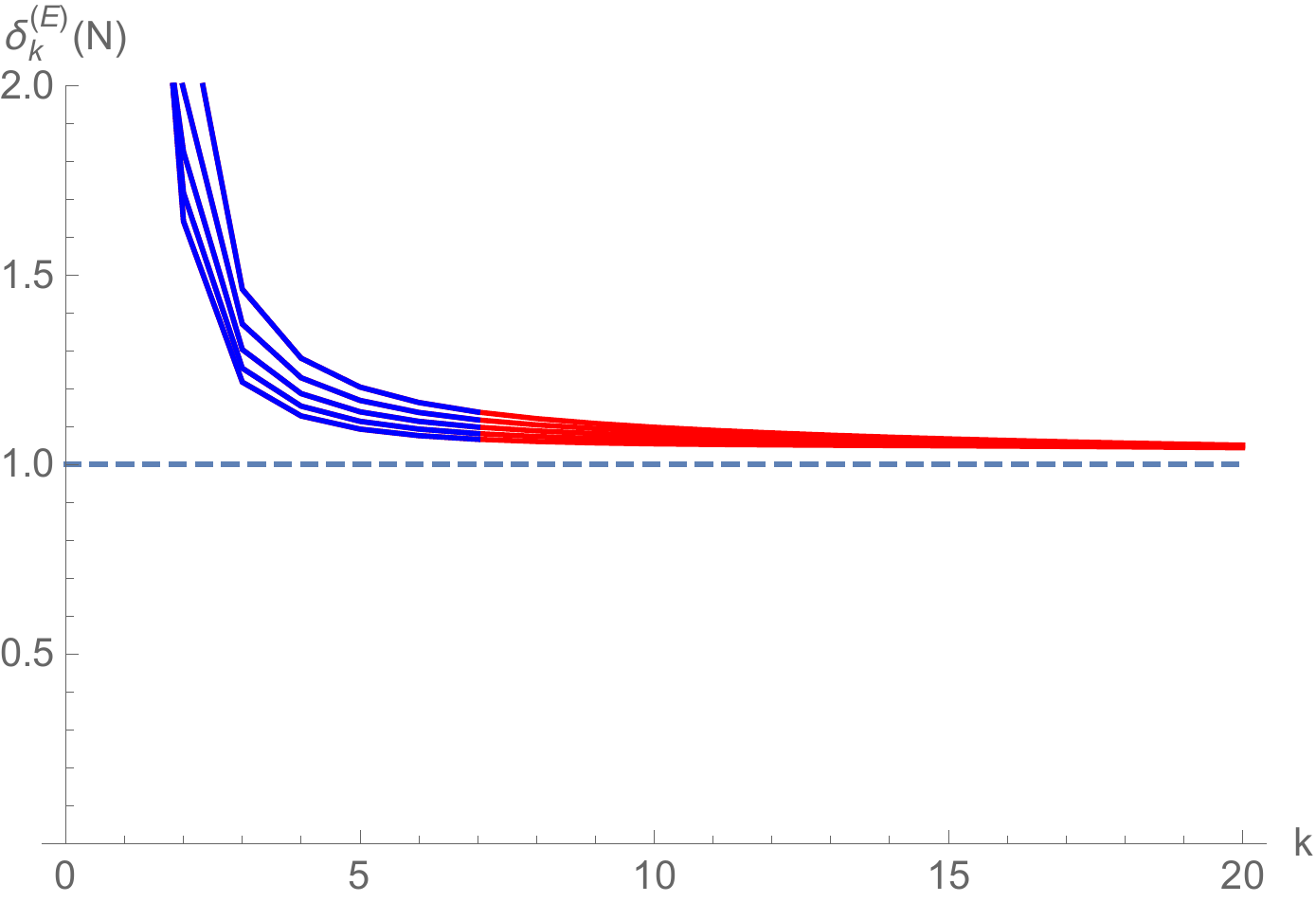}
    \caption{Plots of the ratio $\delta_k^{(E)}(N)$ defined in (\ref{eq:delta-kN-E}) for the $O(N)$ symmetric QM anharmonic oscillator, as a function of perturbative order $k$, and for $N =1,2,\dots,5$. The blue curves are derived from the exact first $7$ coefficients $E_k(N)$, while the red curves continue to 20th order.
Compare with plots of the corresponding combination of coefficients of the perturbative expansion of the beta function (Figure \ref{fig:raw-ratio})  and the epsilon expansion of the correction to scaling exponent $\omega$ (Figure \ref{fig:omega-ratio-plot}).
    }
    \label{fig:energy-delta}
\end{figure}
\begin{figure}[h!]
  \includegraphics[width=7cm]{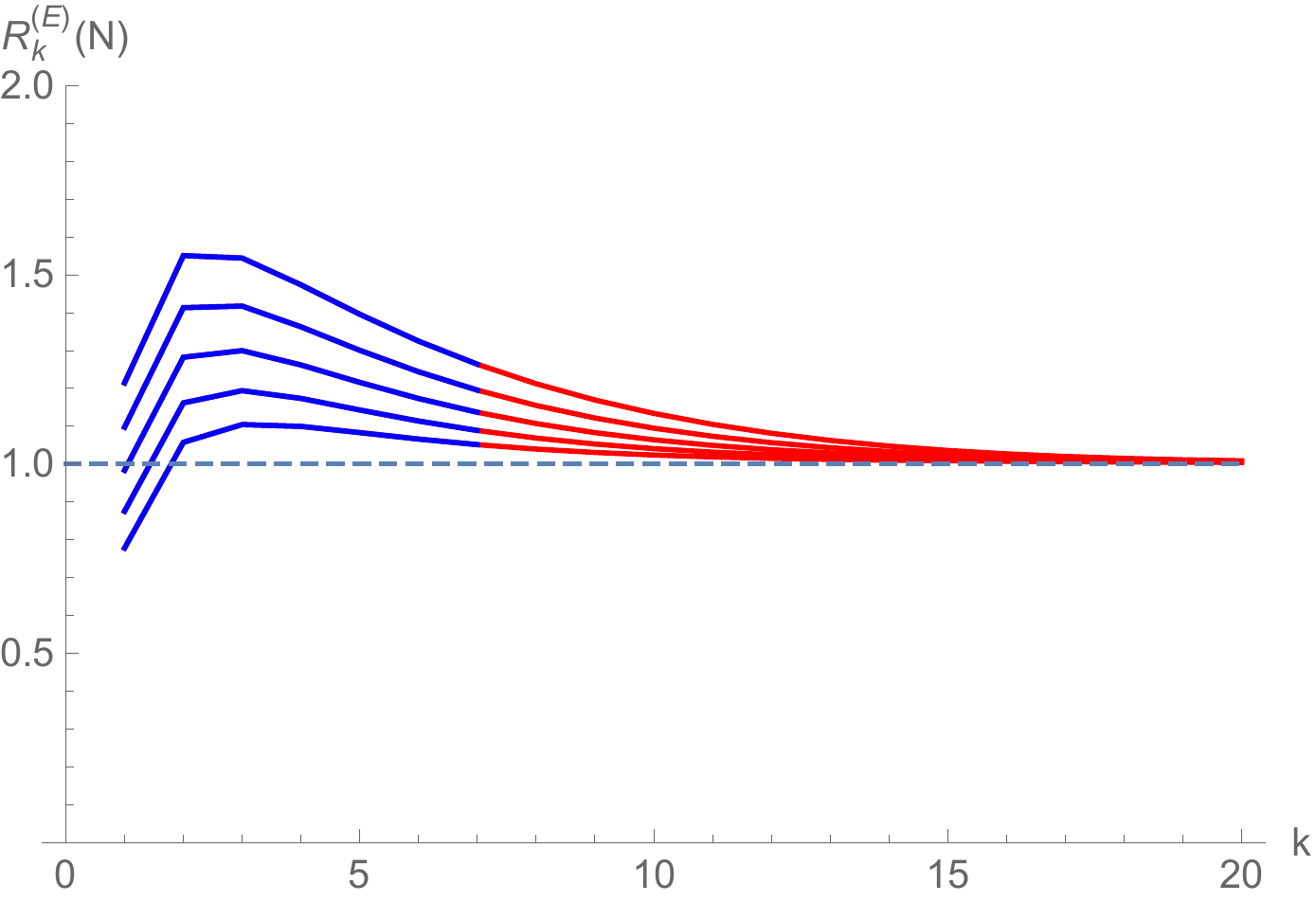}
  \quad
    \includegraphics[width=7cm]{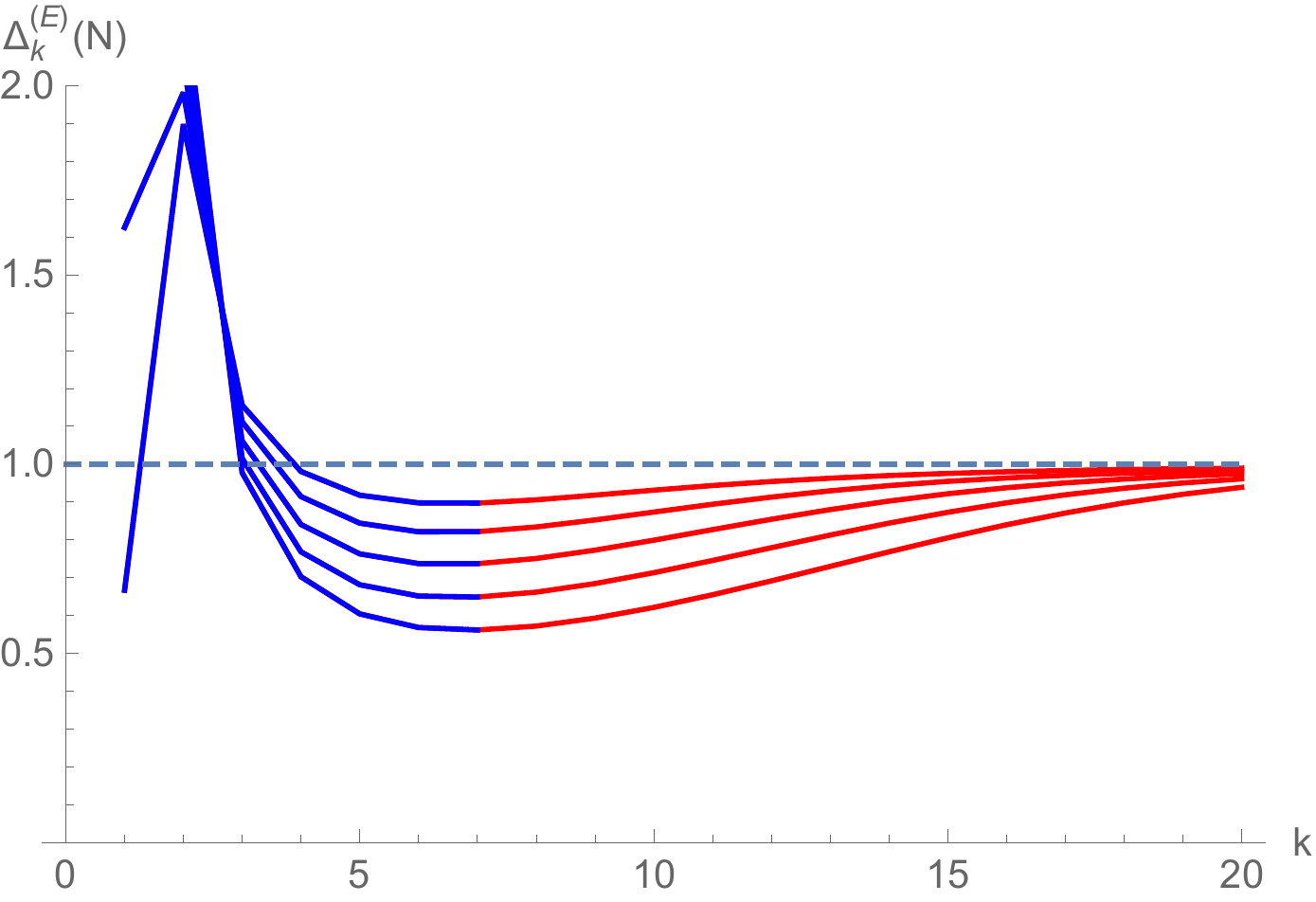}
  \caption{
  Plots of the coefficient combinations ${\mathcal R}_k^{(E)}(N)$ defined in (\ref{eq:R-E}), and $\Delta_k^{(E)}(N)$ defined in (\ref{eq:D-E}),  for the $O(N)$ symmetric QM anharmonic oscillator, as a function of perturbative order $k$, and for $N =1, 2,\dots,5$. The blue curves are derived from the exact first $7$ coefficients $E_k(N)$, while the red curves continue to 20th order.
Compare with plots of the corresponding combinations of coefficients of the perturbative expansion of the beta function (Figures \ref{fig:beta-ratio-inst-plot}
and \ref{fig:beta-delta-inst-plot})  and the epsilon expansion of the correction to scaling exponent $\omega$ (Figure \ref{fig:omega-R-D-plot}).
}
\label{fig:energy-R-D}
\end{figure}

The exact perturbative coefficients $E_k(N)$ can be generated recursively from an algorithm in Appendix A of \cite{Banks:1973ps}. We have computed the first 20, and we define the corresponding combinations from (\ref{eq:r2}), (\ref{eq:ratioratio}) and (\ref{eq:ratio}):
\begin{eqnarray}
\delta_k^{(E)}(N)&:=&\frac{E_{k+1}(N) E_{k-1}(N)}{E_k^2(N)} 
\label{eq:delta-kN-E} \\
{\mathcal R}_k^{(E)}(N)&:=& \frac{-1} { 3\left(k+\frac{N}{2}\right)}   \frac{E_{k+1}(N)}{E_k(N)} 
\label{eq:R-E} \\
\Delta_k^{(E)}(N)&:=&\left(k-1+\frac{N}{2}\right)\left(\frac{E_{k+1}(N) E_{k-1}(N)}{E_k^2(N)} - 1 \right)
\label{eq:D-E}
\end{eqnarray}
These coefficient combinations should each tend to $1$ at large order, for all $N$. See Figures \ref{fig:energy-delta} and \ref{fig:energy-R-D}, which display the combinations derived from the first 7 terms in blue, and those derived from further terms in red. We see that, analogous to the 4 dimensional $\phi^4$ QFT discussed in the previous sections,  the first 7 perturbative orders show similar hints of tending to the correct asymptotic behavior, and this is further improved by higher order terms. But we note that even in this much simpler quantum mechanical model, the true large-order behavior is approached slowly, being much clearer by 20th order than at 11th order, which is the highest order (with approximate estimates) currently available in the 4 dimensional $\phi^4$ QFT.

\section{Conclusions}

We have used the recent high perturbative order exact results of \cite{panzer} and \cite{schnetz} to probe the large order growth of the coefficients of the perturbative expansion of the beta function $\beta(g, N)$ and of the coefficients of the epsilon expansion of the correction to scaling exponent $\omega(\epsilon, N)$ for $O(N)$ symmetric scalar $\phi^4$ theory in 4 dimensions. We suggest that these perturbative results are already showing indications of the generic (factorial $\times$ power) form of large-order growth in (\ref{eq:bwl}). Moreover, the associated large-order growth parameters appear to favor an instanton interpretation rather than a renormalon one, consistent with an argument that renormalization group functions in the MS scheme are not sensitive to renormalons \cite{david,McKane:2018ocs}. This raises interesting questions about the scheme dependence and the observable dependence of renormalons. At present, the diagrammatic understanding of the apparent suppression of renormalons in this case is rather mysterious. It would be interesting to understand better to what extent this is a consequence of the MS renormalization scheme, or the particular renormalization group functions, or both. Nevertheless,  we find it encouraging that exact perturbative computations have matured to the point where they may be on the verge of being able to shed direct light on such questions. Of course, further information about higher orders is still needed to resolve  these issues more conclusively. 
We also note that somewhat related questions have been studied recently using other methods and other QFT models \cite{Sberveglieri:2019ccj,Sberveglieri:2020eko,DiPietro:2021yxb,Marino:2021dzn,Balduf:2021kag}.

 \vspace{.5cm}
\noindent {\bf Acknowledgments} \\
This work is supported in part by the U.S. Department of Energy, Office of High Energy Physics, Award  DE-SC0010339. We thank 
Erik Panzer, Michael Borinsky, Oliver Schnetz,  John Gracey  and Arkady Vainshtein for correspondence and discussions.

\end{document}